# A double-elimination format for a 48-team FIFA World Cup™

César Rennó-Costa[1*]

[1] Instituto Metropole Digital, Federal University of Rio Grande do Norte, Brazil
* contact at cesar@imd.ufrn.br

I present a double-elimination format for the 48-team FIFA World Cup™ that solves many of the concerns raised about the considered formats with mixed round-robin with groups of 3 or 4 teams and single-elimination strategies. Using a quantitative analytics approach, I show that the double-elimination format is fairer, more strategy-proof, and produces more competitive and exciting matches. It solves the problems of possible collusion in the group-of-3 form and the high number of uninteresting games in the group-of-4 format. Using the restrictions of the 2026 FIFA World Cup™ to be held in North America, I demonstrate that the double-elimination format can be implemented in a 35 days window, just a few more than the current format for 32 teams. I discuss how the format can be adapted to the particularities of the host nations in future editions to facilitate attendees' and organizers' planning.

**Introduction**

The FIFA Council unanimously approved that the 2026 FIFA World Cup™ would have 48 teams for the first time [1]. However, the specific cup format with 16 groups of 3 teams raises concern about the possibility of collusion at the last round of the group stage [2]. The constant questioning led FIFA to announce that the Council should consider other cup formats [3]. The alternative format, with 12 groups of four teams, reduces the collusion issue (but not entirely, see [2], [4]–[6]) at the cost of a significant increase in the number of matches and without dealing with some known issues of the current 32-team cup format. The critical points that FIFA was eager to address were the number of uninteresting games at the group stage, a problem that no round-robin format with more than three teams per group can resolve. So, the open question I try to answer is whether an alternative format could be a better fit for the purposes of the FIFA Council [7].

I propose a double-elimination 48-teams cup format for the FIFA World Cup™. Double-elimination formats are group-less, fully knock-out with repechage [8], [9]. The fundamental idea of the format is that the first loss before the semi-finals is not eliminatory, but any team with two losses is eliminated. Matches are scheduled following a swiss tournament model, which produces more competitive games [10], [11]. An undefeated team that reaches the semi-finals can play fewer games before the most crucial championship stage with a whole week of preparation. The one-loss teams have a longer road, which is undoubtedly less frustrating than a single-loss elimination.

Evaluating whether a tournament format would provide a successful FIFA World Cup™ is a multifactorial task that strongly depends on the organizers' vision and the practical limitations [12]. Apart from fairness [13]–[15] – equally skilled teams should have the same chance of winning and the end classification should reflect skillness order – there are other important aspects such as competitiveness [16] – the best teams should face each other -, strategy-proofness [17]–[19] - every team must always have the incentives to win -, injury risks [20]– enough rest time for the athletes to recover between games -, fit to calendar [21] – the tournament should run in a limited window during northern summer -, maximizing attendance and facilitating hospitality [21] – fans should have enough time to organize their travel and be able to foresee where and when its team will play - and drama [22] – the opportunity for historical events that adds to the commercial appeal, maximizing viewership and sponsorship. Although there is some subjectivity in evaluating each of these features, it is possible to apply a more analytical approach [23]–[25].

Here I describe the double-elimination format; compare quantitatively with the group-of-3 and group-of-4 formats using metrics of fairness, competitiveness and tournament duration; propose how it could be implemented at the 2026 FIFA World Cup™ respecting the calendar and athletic requirements of previous editions; and discuss the impact of the format on attendance facilitation and commercial interest.

**The double-elimination format for a 48-teams cup**

The double-elimination 48-teams cup unveils as follows (see Figure 1). In the first round, a draw defines the first 24 matches. These matches are not eliminatory but classificatory. Draws should be resolved by penalty shootouts. The 24 winning teams remain in the main bracket, and the 24 losing teams go to the repechage. The second round has 12 classificatory games in the main bracket between winning teams and 12 eliminatory games in the repechage route between the losing teams. At the end of this stage, 12 teams leave the championship with two losses. In the third round, I have six classificatory games between the winning teams of the main bracket and 12 eliminatory games where the 12 winners of the previous repechage stage face the 12 losers of the main bracket games. Once again, 12 teams are eliminated with two losses and one win.

In the next stage, two out of the 18 teams with one loss and two wins are allowed back to the main bracket to balance the confrontation scheme. Standard group classification rules apply. Therefore, it is very likely that these teams' losses were due to a draw with penalty shootouts. Now, I have eight teams in the main bracket and 16 teams in the repechage. The four undefeated winners are organized into two matches that define the two undefeated teams classified into the semi-finals.

Eight out of the 16 teams are eliminated in the repechage with two wins and two losses. The eight winners join four losers of the main bracket to play six new eliminatory games. Six out of the twelve are eliminated with three wins and two losses, and the remaining six join the two losers of the last main brackets games. The remaining eight teams play a standard knock-out eliminatory bracket until there are two remaining that will face the undefeated teams in the semi-finals.

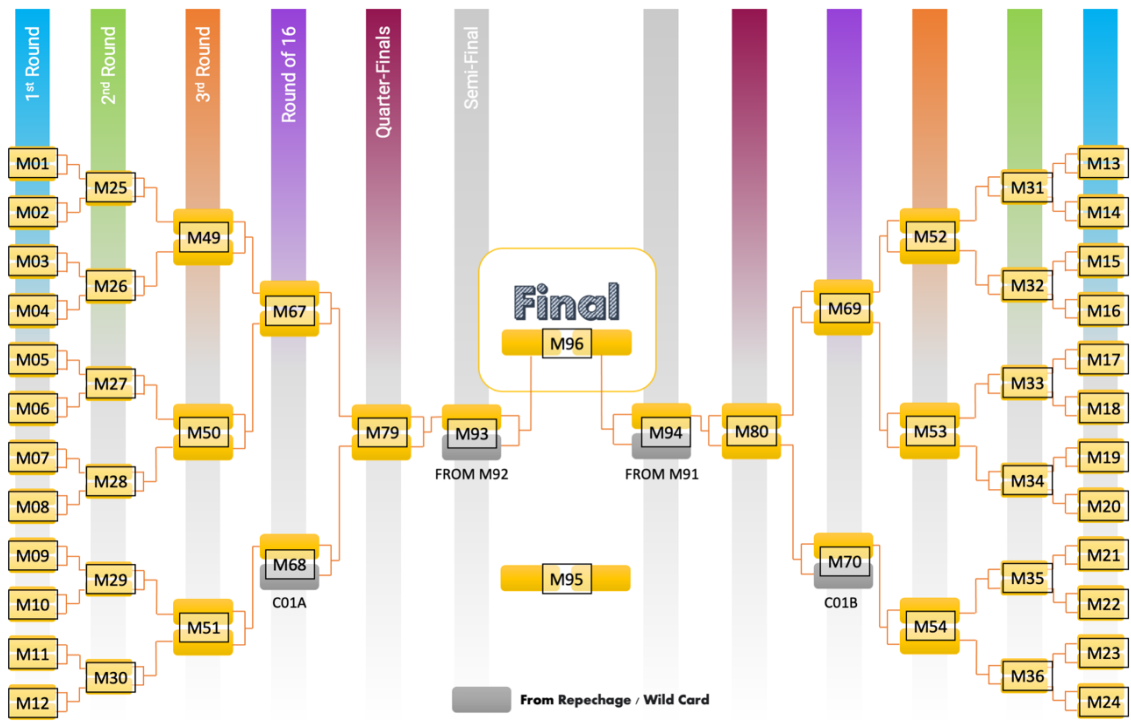
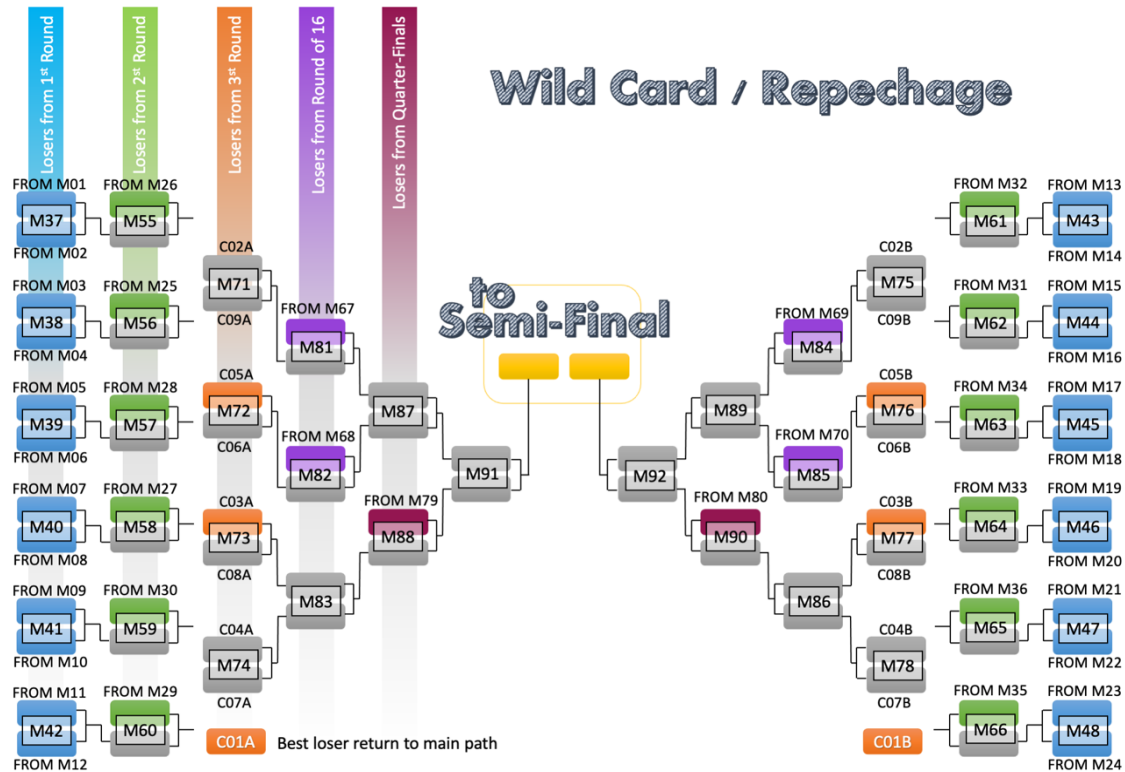

*Figure 1 Double-elimination format for a 48-teams FIFA World Cup™. (top) Main bracket for the undefeated teams to reach the championship. Silver boxes indicate entry points for teams in the repechage. (bottom) Repechage bracket that leads to a return to the main bracket. Losing teams from the main bracket enter the repechage at different levels indicated in as colored boxes.*

The semi-finals are played as in the current format. This double-elimination format guarantees two undefeated teams that played five games in the semi-finals and two squads with one loss and six wins (with a total of seven games). As a result, an undefeated team can win the championship after seven games, just as in the current 32-teams format. A team with one loss will require nine games to become champion.

The proposed format's main advantage is that every single game is decisive. It ensures that one team will only be eliminated after a defeat and that there is no elimination after a victory. As all games matter, the format avoid the common situation where strong teams use alternative players in a game that has little impact on their tournament trajectory. This situation has led to events in the past where a team was eliminated after a historic win because, in the other game, one of the teams did not use its forces at full. Also, it avoids the matches in which both teams are already eliminated. If a team is playing, it still has a chance to win the cup.

The second advantage - which is not so evident - is that elimination for the undefeated is only possible in the semi-final. This aspect allows a strong and undefeated team that reaches the quarterfinals, for instance, to return to the competition after a frustrating loss on a more difficult path. As a result, the format solve the issue of having very seldom games in the last stages and a very fast elimination of strong teams. The practical result is that one can expect more matches between top-ranked teams with this format (see below).

The main disadvantage of the format is that depending on the win or loss of a team in the main bracket, the team has different paths to follow that can be difficult for the fans to follow. The organization can handle the scheduling challenges by planning the winning and losing matches at nearby venues or even the same stadium, as in Qatar. The other main issue is that a champion team might require nine games to become a champion, which might provoke significant physical stress on players. Stress can be mitigated by only allowing extra time in the semi-final and final stages.

**The other formats**

For this analysis, I consider the two tournament formats considered by the FIFA Council: the 18 groups of 3 teams and the 12 groups of 4 teams. In the group-of-3 format, each group has three games between each of the teams. The first two teams classify for a Round of 32 knockout tournament format with single elimination. In the group-of-4 format, each group has six games between the teams in three rounds. In the format I consider here, the two best teams of each group qualify for the Round of 32 knockout tournament as well the three best third places of each side of the brackets (first 24 teams and the last 24 teams). The knockout proceeds as a single elimination format.

The key advantages of these modes is that there is no bye and the only path to the championship is to play seven or eight matches for the group-of-3 and group-of-4, respectively. The key disadvantage of the group-of-3 format is the risk of collusion, as one team will rest in the 3rd round of the first stage. Another issue is the disbalance of rest days between teams considering the Best of 32 stage. The group-of-4 has the main issue of the number of matches, the number of uninteresting games and the fact that it only partially eliminates the risk of collution.

**Analytic evaluation of cup formats**

I used a Monte Carlo strategy (see below) to quantify the main differences between the double-elimination proposal and the group-of-3 and group-of-4 formats. The method consists of developing multiple cup simulations and evaluating the final classification. The simulation engine is a computational model that accurately predicts a team's chances to win or draw based on the differences in the teams' FIFA ranks. I used information from previous FIFA World Cups™ to evaluate how each team performed based on their FIFA rank position.

The main requirement of a tournament format is to be fair. I define fairness as a championship whose final classification reflects the skillness order of the participants, with more importance on the final position of the best-qualified teams. I evaluated each of the championship formats using a fairness metric that measures the actual position of the team and the position expected for its relative skill level. The metric weights distance with higher values for the first positions and lower weights for the last positions. After the simulation of 5000 tournaments of each format, I found that the double-elimination has a slightly higher probability of producing fairer competitions than the group-of-3 and group-of-4 formats (Figure 2).

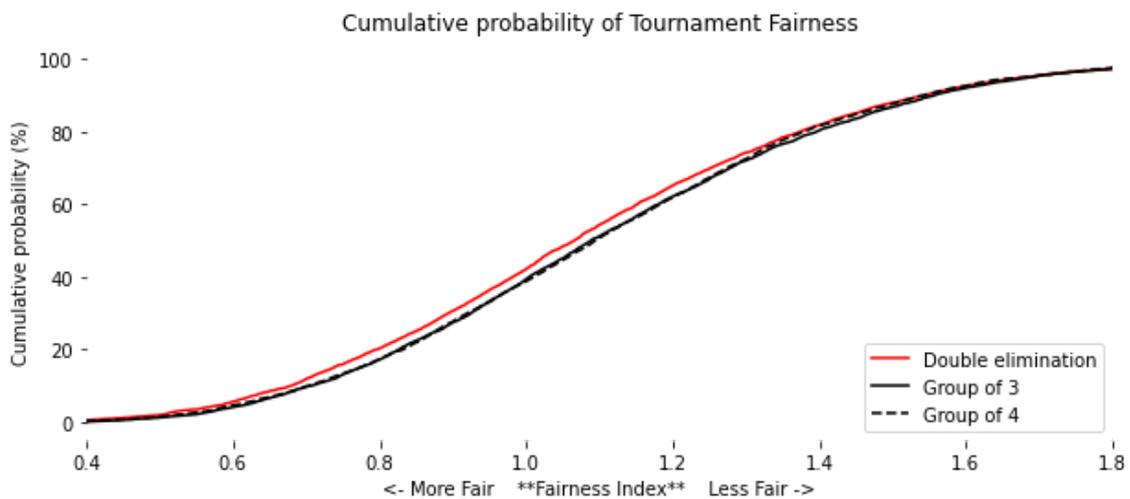

*Figure 2 Double-elimination format provides fairer tournaments than other formats. The cumulative probability of a tournament to have a fairness index of at least the value indicated. The double-elimination format has a slightly higher chance of producing tournaments with lower Fairness Index.*

Another key feature of a championship format is the quality of the games it produces. I established a Rank Index to qualify each match by the FIFA Rank of each team. Matches with two high-ranked teams have a Rank Index next to 100, whereas a game with two low-ranked teams has a Rank Index next to zero. To evaluate how each format favors high-ranked matches, I simulated 5000 cups of each format and compared them with a random distribution of matches between the teams that would appear in the cup (Figure 3). I found that the three formats favor high-ranked matches, which is unsurprising because high-ranked teams have a higher chance of advancing in every format. However, I also found that the Repechage format favors the occurrence of matches with a Rank Index above 75 points. These are games where both teams are among the TOP20 ranked teams.

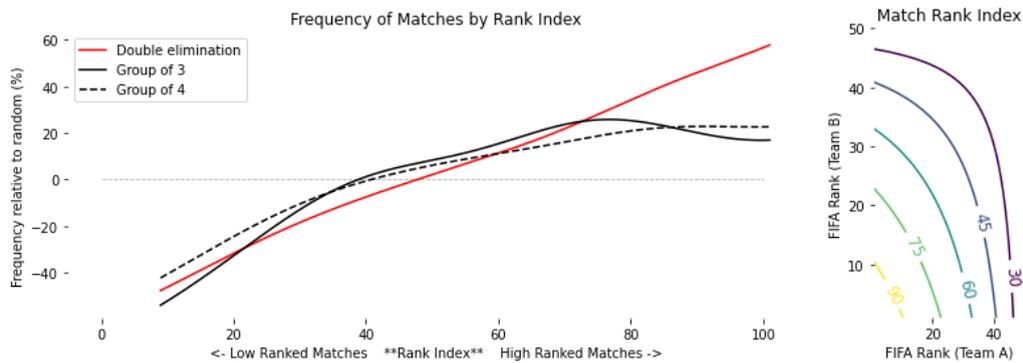

*Figure 3 Double-elimination format provides more high-profile games than other formats. (left) The expected frequency of games based on their Rank Index, relative to a random assignment of matches. Shown for the double-elimination, group-of-3 and group-of-4 formats. (Right) Indication of the Rank Index values for matches between two teams (A and B) and their respective FIFA Rank position.*

The second metric of match quality relates to the competitiveness level. A competitive game is a game where the skill difference between the competitors is low. Using the metric of Rank Distance, i.e., the average distance of FIFA rank from the teams in each match, I found that the double-elimination format has a higher probability of producing competitive matches than the other two formats in a simulation of 2500 cups (Figure 4).

Although the double-elimination format provides more exciting games, would this come with the cost of a higher number of low-ranked confrontations as well? Indeed, the group-of-3 format is the most compact option, with 80 matches. The group-of-4 format significantly increases the number of matches to a total of 104 games, representing an increase of 30%. Our double-elimination format requires 96 games, while still a 20% increase if compared with the group-of-3 format; it means about 10% fewer games than the group-of-4 format.

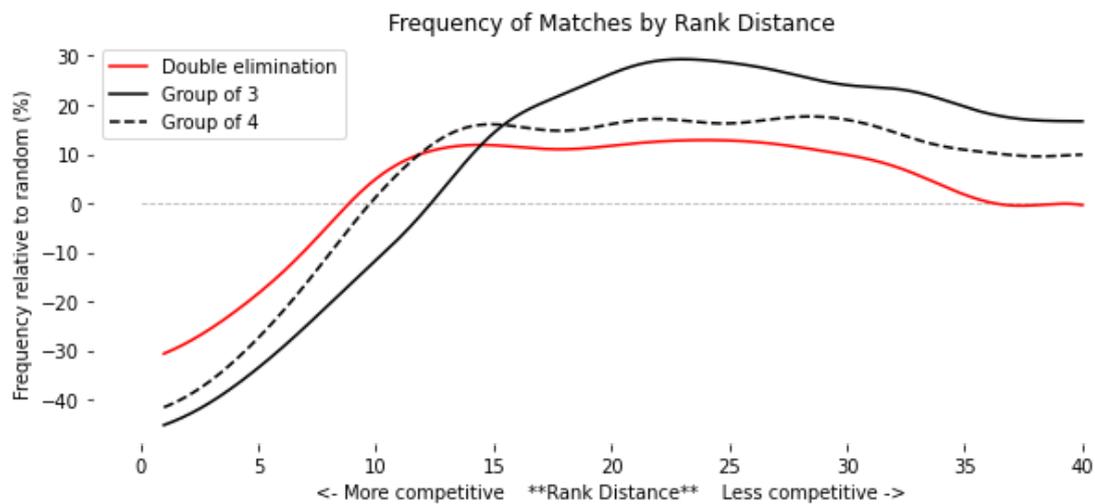

*Figure 4 Double-elimination format provides more competitive games than other formats. (left) The expected frequency of games based on their Rank Distance, relative to a random assignment of matches. Shown for the double-elimination, group-of-3 and group-of-4 formats. (Right) Indication of the Rank Distance values for matches between two teams (A and B) and their respective FIFA Rank position.*

While the double-elimination format represents an increased number of matches, these games are mainly of high interest (Figure 5). The double-elimination format provides an average increase of 62% on high-interest games where both teams are within the TOP8 teams if compared with the group-of-3 format and 48% with the group-of-4 format. On special-interest games, where one of the teams is among the TOP8, the double-elimination format provides 19% more games than the group-of-3 format and about 5% fewer games than the group-of-4 format. Considering regular games, where none of the teams is a TOP8, the group-less format provides 18% more games than the group-of-3 and 11% fewer games than the group-of-4 format. Therefore, the repechage format requires more games than group-of-3, but these are high-interest games. The alternative group-of-4 format provides an even higher increase in the number of matches but without a strong impact on the number of games that call the public's attention. Moreover, as all games are classificatory or eliminatory, even those without top-ranked teams will have a strong public appeal.

The increase in the number of games does not significantly impact the number of days required to run the cup (Figure 6). The current 32-team format runs for 29 days in Qatar 2022, 32 days in Russia 2018, and 32 days in Brazil 2014. The great advantage of the group-of-3 format is that it could be played in the same 32 days, considering a maximum of 4 games per day and a rest time between games of 4 days. None of the options would fit in a 32-days window with these conditions. The group-of-4 and the repechage - with a few 3-day-rest games (as in Qatar 2022) - formats could fit in a 39-days window (one extra week compared to the current format). Changing the number of matches allowed in a single day has an important impact on the duration of the cup. With five games per day, the double-elimination format could be played in 38 days. Increasing the number of matches is only necessary on the first few days of the cup.

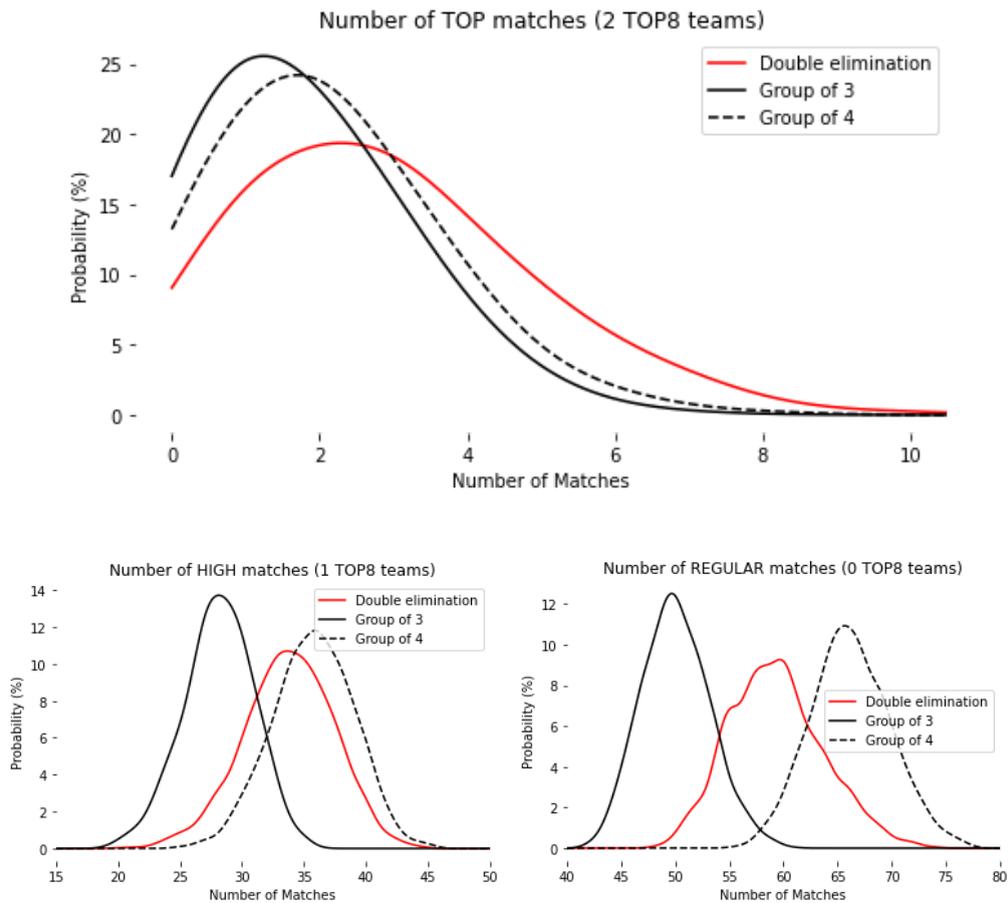

*Figure 5 Double-elimination format increases the number of interesting matches but not much of the regular matches. (Top) Probability of having a specific number of matches between two TOP8 teams, shown for the three formats. (Bottom left and right). Same as above, but for matches with one or none TOP8 teams.*

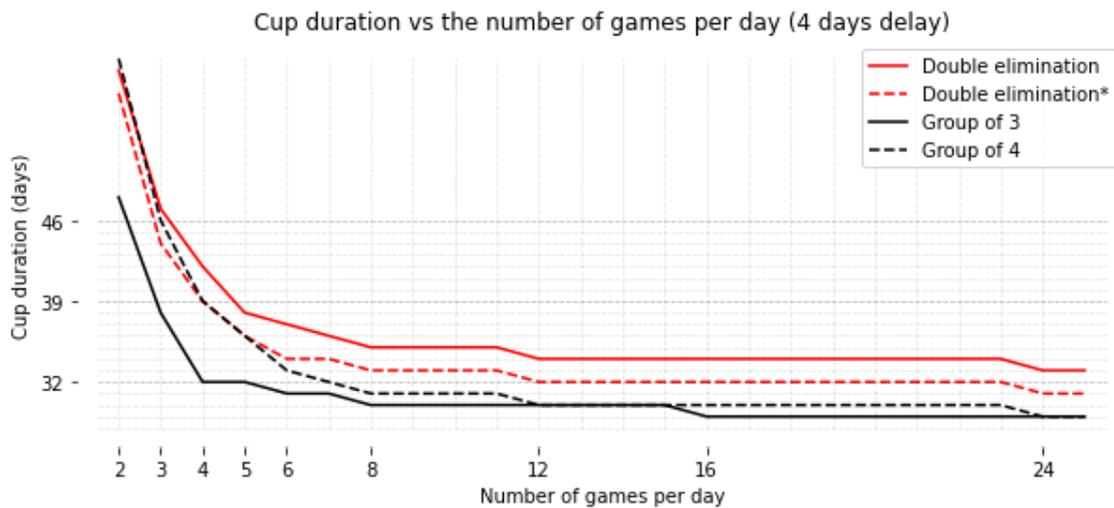

*Figure 6 Double-elimination format demands more days to organize the cup than other formats but increasing the number of matches in a single day, makes it require just a couple of extra days. The graph shows the minimum number of days to organize the cup for each format considering the minimum of 4 days between games of the same team and a variable number of matches on the same day. double-elimination is also shown with the possibility of 3 days between repechage matches.*

## 2026 FIFA World Cup™ at USA, Canada, and Mexico

The 2026 FIFA World Cup™ will happen in June and July in North America, with matches at 16 host cities distributed in the USA, Canada, and Mexico (Figure 7). I present one tentative schedule with the FIFA World Cup™ starting on June 15th and finishing on July 19th (Figure 8). I consider a total of six games per day during the first days of the tournament. The unusual placement of the host cities in multiple time zones allows many time slots for prime-time broadcasting in America, Europe, and Asia.

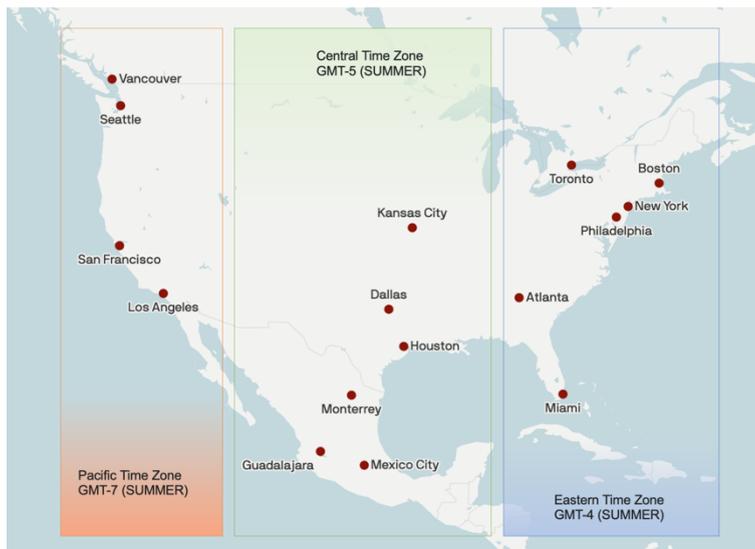

Figure 7 The 16 host cities for the 2026 FIFA World Cup span for 4 different time zones. Matches will be held at UTC-7, UTC-5, and UTC-4.

Figure 8 A tentative schedule for the 2026 FIFA World Cup™ in the double-elimination format. For the brackets, check Figure 1.

The high number of host cities in three countries spanning four time zones presents a challenge for a double-elimination cup format, as an unknown team path will be difficult to plan for the attendees and organizers. Traveling between some host cities can require up to 6 hours of flight time, and crossing borders can be bothersome for attendees requiring a VISA. Organizers planned to use Canada and Mexico host cities only during the first two matches in the group-of-3 format to mitigate the traveling issue above. In the double-elimination format, Canada and Mexico host cities should be used for the first three rounds, guaranteeing that both teams would play on their home turf (or at least close to the border in the case of Canada with cities like Seattle or Boston). Interestingly, the bracket allows multiple options for each home team, which helps to include these teams in the draws at specific positions in the brackets. Also, the table clearly distinguishes between the first and left sides of the brackets, which only mix after the third round. This makes organizing the host cities easier, so fans and teams have fewer travels during the games. The schedule also provides a week-long rest for the two undefeated teams before the finals games.

In this format, the winning teams know their path to the final from the first game. Changes occur when the team loses. However, the schedule is organized so that changes in the bracket lead to games in nearby cities.

**Discussion**

I presented an alternative format for the FIFA World Cup™ with 48 teams based on double-elimination and swiss tournament concepts. I show that the proposed format is feasible in 35 days - just a few days more than the current formats of a 32-teams cup - and is fairer, more strategy-prof, more competitive, and more interesting than the other formats.

Other formats could also be considered, such as having an entire knock-out tournament. The knock-out stage of the FIFA World Cup™ is fully packed with emotions as, by definition, every game is decisive. However, the eliminatory nature of these games creates a different issue: the team that loses returns home. What if a team loses in the very first game? That disappoints many fans that flock to the FIFA World Cup™ host nations every four years. Group stages and the double-elimination format ensure that teams with an early loss can recover, as seen in the 2010 and 2022 FIFA World Cups™, in which the champions lost their first game in the competition. Indeed, one of the proposals for the 48-teams format included a first knock-out stage in which 12 teams would leave the tournament after a single match. The FIFA Council promptly rejected this format.

Other less traditional formats exist, including mixtures of groups and single elimination, purely swiss system tournaments, league format with ongoing elimination, and others. I did not consider these options in the analysis as these designs demand some effort to adapt to the rigid requirements of a FIFA World Cup™.

One main advantage of the double-elimination format is to take advantage of the current trend of organizing the FIFA World Cup™ in multiple countries and venues. For 2030 there are already numerous multi-country bids such as the Argentina-Uruguay-Paraguay-Chile bid [27] and the first three continents, Egypt–Greece–Saudi Arabia bid [28]. The double-elimination format allows the organization of the brackets into small nucleus and, as shown for the 2026 tournament, increases the distance as the team advances in the championship. In the schema presented here, there are multiple slots for the host teams to guarantee to play on their home turf for at least the three first rounds. If the organization desires it, extending this guarantee for the next rounds is possible.

The FIFA Council has been very innovative throughout the years in the FIFA World Cup™, introducing technologies such as VAR, semi automatic control of offside and ball position. The esports tournaments of FIFAe already apply a double-elimination format with great success [29], [30]. Why not change the tournament design to a more dynamic one, which will undoubtedly lead to a more emotional and competitive event? The fans will surely appreciate it.

**Methodology**

**Simulations**

All simulations were done using Python code with use of Numpy library. The executable code was organized as a Jupyter Notebook, which is openly available at GitHub.

**The result model**

To support the Monte Carlo engine of the cup simulations, I developed a stochastic decision model based on the FIFA Rank of each of the confronting teams. The model defines randomly the number of goals of each of the teams based on a Poisson distribution with a variable Lambda parameter set as a function of the teams FIFA Rank (here stored at the team_rank variable):

```python
number_of_goals = np.random.poisson(1.5+0.7*(1-2*(min([team_rank,50])/50)))
```

For each match, I randomly assigned the number of goals of each team and computed the result as normal. Penalty shootouts winners were defined randomly with equal chance independently of the FIFA Rank. The model accurately fits the observable data from the FIFA World Cups™ between 1994 and 2014 (Figure 9).

**Fairness Index**

To compute the Fairness Index (FI), I compute the final classification of the championship as a sequential list of all participants. As a reference, I establish a skill order list relative to the FIFA Rank proxy. The FI is computed as follow:

```python
fairness_index = sum([(abs(position-rank)/48)*((1-(rank/48))**gamma) for position,rank in enumerate(cup_classification)])
```

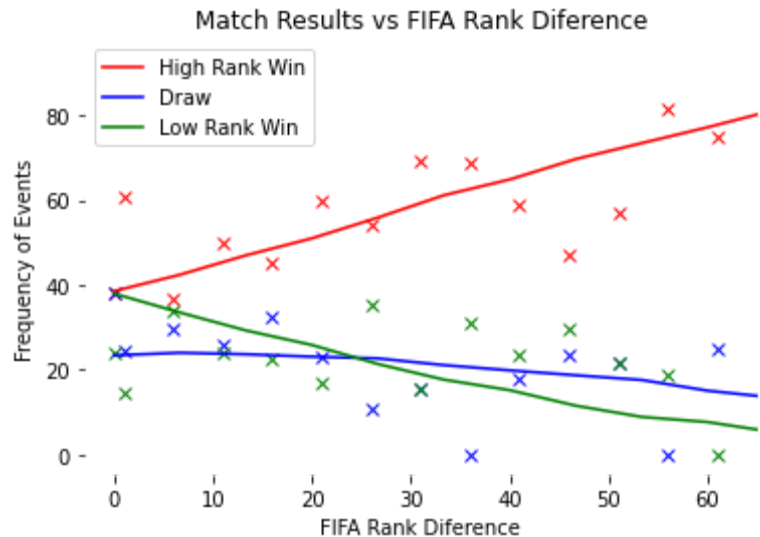

*Figure 9 Match results from the Poisson-based Monte Carlo engine accurately reproduce the distribution of wins, draws and loses, based on the FIFA rank difference of the competitors. Crossed are the grouped values of actual matches on the FIFA World Cups™ from 1994 to 2014. Lines indicate the output frequency of events of the model.*

**Rank Index**

The Rank Index (RI) is computed for each match of the tournament as follow:

```
norm_team_rank_home = 1.0-float(min([team_rank_home,50])-1)/50.0
norm_team_rank_away = 1.0-float(min([team_rank_away,50])-1)/50.0
match_rank_index = np.sqrt(norm_team_rank_home*norma_team_rank_away)
```

**Rank Distance**

The Rank Distance (RD) is computed for each match of the tournament as follow:

```
rank_distance = abs(team_rank_home-team_rank_away)
```

**Cup duration**

The cup duration is computed based on two parameters: the maximum number of matches per day and the minimum days of rest for each match considering the previous games. For each format I established the dependencies of each match. The games were allocated in order following the two requirements. When a day is full, the game is allocated to the next available matchday. The duration of the cup was computed as the number of days between the first and the last matches.

**Acknowledgement**

I thank Vitor Lopes dos Santos and Luana Ferraz Alvarenga for the valuable discussions.